# A Direct Power Controlled and Series Compensated EHV Transmission Line

Andrew Dodson, *IEEE Student Member,* University of Arkansas, amdodson@uark.edu

Roy McCann*, IEEE Member*, University of Arkansas, rmccann@uark.edu

**Main Contact**
Email: amdodson@uark.edu
Department of Electrical Engineering
3217 Bell Engineering Center
Fayetteville, AR 72701, USA

**Abstract –** This paper presents the design and analysis of a compensation method with application to a 345 kV 480 MVA three-phase transmission line. The compensator system includes a series injected voltage source converter that minimizes the resonance effects of capacitor line reactance. This creates an ability to compensate for the effects of subsynchronous resonance and thereby increase line loadability and control real and reactive power flows. The granularity of power flow control and simultaneous stabilization is achieved by the method of direct decoupled power control (DPC). The design process is detailed with respect to optimal response characteristics considering variations of line parameters, realistic transformer impedances, and maximum ramp response rates. Line effects are demonstrated in a PLECS model in MATLAB, and compensation control system functionality is verified. A case study is provided of a 345 kV transmission line from an EMTP simulation in PSCAD that accounts for distributed parameter effects that are encountered in physical EHV transmission lines. This demonstrates the improvement in stability to power system transients as well as damping of power system oscillations.

## I. Introduction

There is continuing to be increased adoption of wind and solar energy production in the central region of North America. In general, these are intermittent in nature and therefore are not dispatchable. Moreover, these renewable sources are most abundant in rural areas that are distant from primary electrical loads. As a result, there is a large portion of electric power that is transmitted over long (in excess of 250 km) EHV transmission lines. This creates an operational hazard in that series capacitors may be required in order to stabilize power flows. As has been observed since the 1970's, the use of fixed capacitors may lead to subsynchronous resonance effects [5]. For this research, a compensation technique using a static synchronous series compensator (SSSC) is used for overcoming the effects of subsynchronous resonance effects. For transmission under 250 km, a simple RL model can be used, but for longer lines, a split-$\pi$ model is a good approximation of the distributed parameter effects of line transmission lines. The split-$\pi$ model is used for design and analysis purposes in this research and is shown in Fig. 1. The validation of the design is based on a complete distributed parameter model (i.e., a line modeled by partial differential equations) through the use of an Electromagnetic Transients Program (EMPT) solver [13]. This approach allows for understanding the underlying principles for compensation of long transmission line effects, which results in specific design methods that can be employed by transmission system planners. The method also fully comprehends the underlying physics and mathematical principles of transmission line transient and dynamic effects.

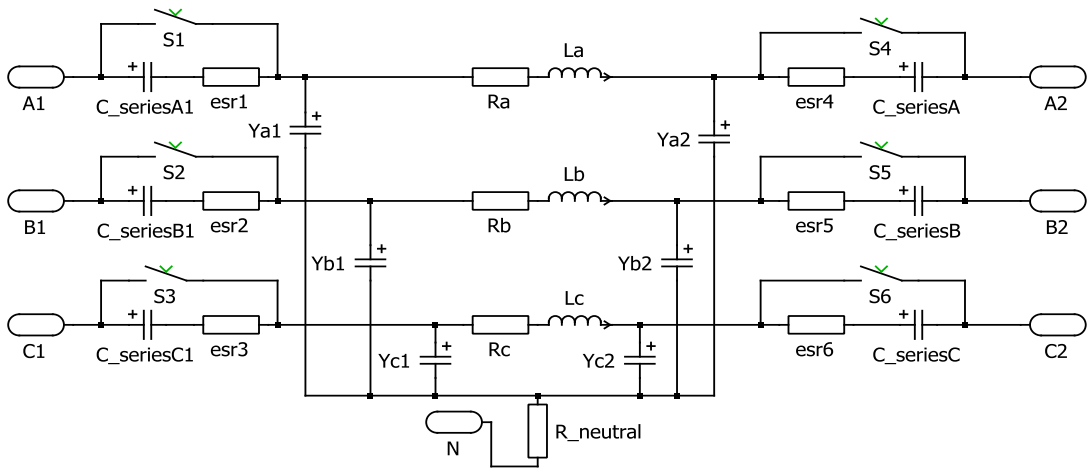

Fig. 1. Series compensated three-phase transmission line split-$\pi$ model.

## II. System Description

Long lines can cause issues with voltage sags or swells in the steady state; while power transients can cause ringing between the line series inductance and shunt admittance. In addition, generators separated by long lines must maintain synchronous rotation speeds in phase with their respective bus voltages. In order to improve line loadability, series capacitance is often inserted to counteract the line inductance. Typically this capacitance is subdivided and located equidistant along the line. The theoretical limit on line power flow occurs when the angle from the sending to receiving bus is 90 degrees, giving $P_{max} = V_S * V_R/X'$. Compensating with a capacitance of $C = \frac{N\%}{2*\omega^2*L}$ gives an increased power flow of $\frac{1}{1-N}\%$ which is extremely valuable from a transmission and distribution upgrade deferral perspective. Upper bounds to the size of series capacitive compensation are due to the size and weight considerations of installing equipment on high power line towers, typically limiting to 30% or less.

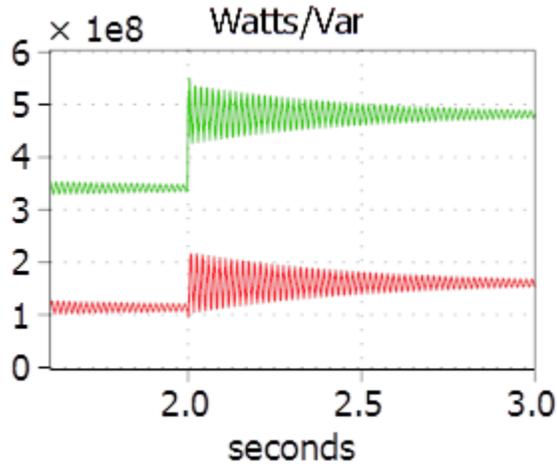
Fig. 2a: Effect of line compensation.

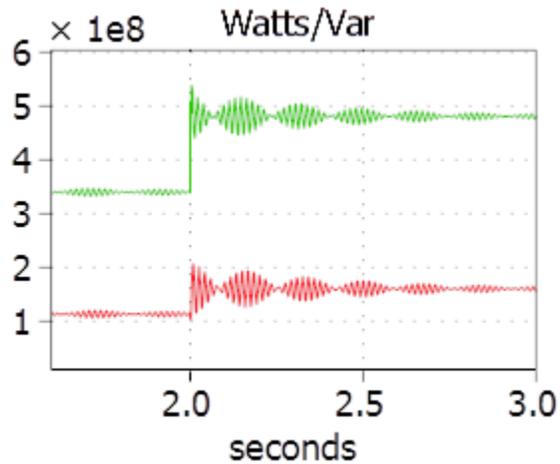
Fig. 2b: Effect of increased line compensation.

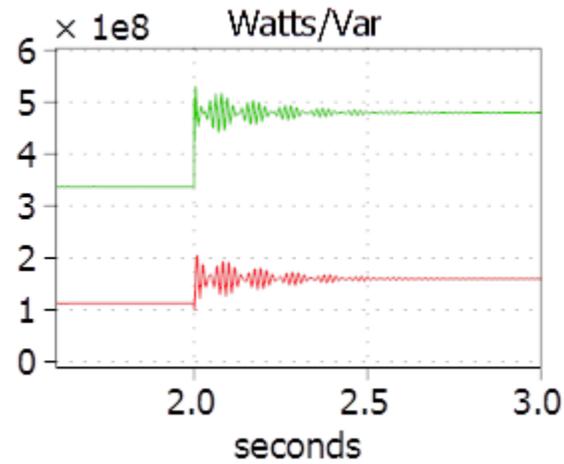
Fig. 2c: Effect of increased line compensation.

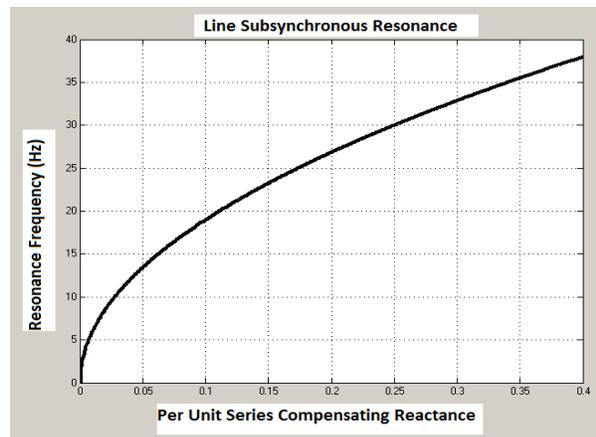
Fig. 2d: Effect of compensation on resonance frequency.

Fig. 2. Series Capacitive Compensation P and Q   (a) 0%  (b) 5%   (c) 15%   (d) SSR vs. $N_{p.u.}$

The main disadvantage of this form of compensation is that a subsynchronous resonance (SSR) introduced at $f_{res} = \sqrt{N} * 60\ Hz$. These resonances are dangerously close to the natural frequencies of mechanical oscillations in power generation equipment. Power transients cause rotating machinery to undergo accelerations due to mismatch of instantaneous electrical and mechanical power. This fatigues the generator shaft with damaging torque swings and can lead to mechanical failure or loss of synchronicity. Figure 2 demonstrates the increase in line stability due to series compensation. In each case the line is stepped from 360 MVA to 480 MVA at a power factor of 0.8. The ringing of the line becomes bounded by the SSR oscillatory mode, which increases in frequency and attenuation as per unit compensation increases.

In a networked transmission grid, such as Fig. 3, there are typically many paths from sources to sinks of energy. Different paths through the network will have differing impedances dependent on line length and voltage, as well as limitations on maximum current flow due to thermal or stability considerations. For lines longer than 250 km, the stability limit on power transfer dominates the thermal limit. Low impedance paths will carry the most current for a given voltage difference between a sending and receiving bus. Thus the transmission grid's ability to transfer energy is limited by those conduction paths with the lowest impedance. Shunt compensation is concerned with power quality issues of a single bus, cannot be used to solve the problem of equal current sharing on separate lines [8]. A series compensator with the ability to vary the apparent line impedance is required for equalizing power flows in parallel or more complexly networked grids. A system that addresses these two issues simultaneously would be of significant value to independent system operators. DPC utilizes the Clarke transformation (or Park and a phase locked loop to obtain instantaneous phase angle), to decompose a set of three phase vectors into the quadrature components.

### III. Dynamics of the Capacitively Compensated Line

A simplified RLC model is used as the basis of determining the line dynamics as in Fig. 3 [5]. This ignores the shunt admittance of the line, but this is acceptable as the dynamics of the SSR and steady state power flow are of primary concern. The voltage source, $V_{con}$, would be implemented via a battery energy storage system (BESS) [7], an impedance sourced converter (ZSC), and a single turn transformer (STT) [6]. The STT ensures appropriate device voltages and limits fault current. The ZSC topology provides an interface for the variable DC voltage of the BESS as well as eliminating the failure mode due to capacitor shorting through one of the bridge legs of the converter. This system provides the high reliability required of grid connected power electronics systems.

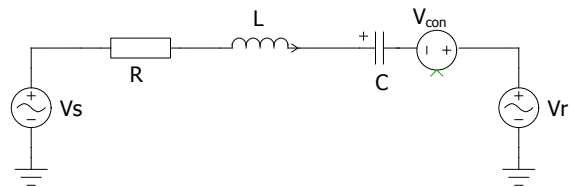

Fig. 3. RLC line model.

Combining $V = V_S - V_R + V_{CON}$, the line dynamics are:

$$L * \dot{I} = -V_C - R * I + V$$
$$C * \dot{V}_C = I \qquad (1)$$

Applying the quadrature property of the Clarke transform [3] allows elimination of the capacitor voltage as a state variable,

$$V_\alpha(t) = |V|\sin(\omega t)$$
$$V_\beta(t) = -|V|\cos(\omega t) \qquad (2)$$

$$\dot{V}_\alpha = -\omega V_\beta = I_\alpha/C$$
$$\dot{V}_\beta = \omega V_\alpha = I_\beta/C \qquad (3)$$

This reduces the line current dynamics to

$$\dot{I}_\alpha = \frac{1}{L}\left(-\frac{1}{\omega C}I_\beta - RI_\alpha + V_\alpha\right)$$
$$\dot{I}_\beta = \frac{1}{L}\left(\frac{1}{\omega C}I_\alpha - RI_\beta + V_\beta\right) \qquad (4)$$

It is now possible to obtain transfer functions from line voltages to line currents as

$$\begin{bmatrix}I_\alpha\\I_\beta\end{bmatrix} = \begin{bmatrix}G_{\alpha\alpha} & G_{\alpha\beta}\\G_{\beta\alpha} & G_{\beta\beta}\end{bmatrix}\begin{bmatrix}V_{S,\alpha} - V_{R,\alpha} + V_{CON,\alpha}\\V_{S,\beta} - V_{R,\beta} + V_{CON,\beta}\end{bmatrix} \qquad (5)$$

$$G_{\alpha\alpha} = G_{\beta\beta} = \frac{C^2\omega^2(R + Ls)}{C^2L^2s^2\omega^2 + 2C^2LRs\omega^2 + C^2R^2\omega^2 + 1} \qquad (6)$$

$$G_{\alpha\beta} = -G_{\beta\alpha} = -\frac{C\omega}{(L^2s^2\omega^2 + 2LRs\omega^2 + R^2\omega^2)C^2 + 1} \qquad (7)$$

**IV. Direct Power Control of Series Injected Voltage**

Direct power control (DPC) methods have been developed for electric power transmission systems ([1],[2]). Damping of subsynchronous resonance has previously considered shunt compensation circuits [4]. This research explores the DPC technique to long transmission lines using series compensation. In the $\alpha\beta$-frame, the utility of the Clarke transform [3] becomes apparent when calculating complex power,

$$\begin{bmatrix}P\\Q\end{bmatrix} = -\frac{3}{2}\begin{bmatrix}V_{S,\alpha}I_\alpha + V_{S,\beta}I_\beta\\V_{S,\beta}I_\alpha - V_{S,\alpha}I_\beta\end{bmatrix}, \qquad (8)$$

where the negative sign indicates power flowing from the sending end of the line. Substituting in the appropriate transfer functions, the line power flow can be calculated directly from the voltages as

$$\begin{bmatrix}P\\Q\end{bmatrix} = -\frac{3}{2}\begin{bmatrix}V_{S,\alpha}(G_{\alpha\alpha}V_\alpha + G_{\alpha\beta}V_\beta) + V_{S,\beta}(G_{\beta\alpha}V_\alpha + G_{\beta\beta}V_\beta)\\V_{S,\beta}(G_{\alpha\alpha}V_\alpha + G_{\alpha\beta}V_\beta) - V_{S,\alpha}(G_{\beta\alpha}V_\alpha + G_{\beta\beta}V_\beta)\end{bmatrix}. \qquad (9)$$

Finally, the power dynamics are obtained, with appropriate substitutions for P and Q, as

$$\frac{dP}{dt} = -\frac{3}{2L}[(V_{S,\alpha}{}^2 + V_{S,\beta}{}^2)-(V_{S,\alpha}V_{R,\alpha} - V_{S,\beta}V_{R,\beta})+(V_{S,\alpha}V_{CON,\alpha} - V_{S,\beta}V_{CON,\beta})] - Q\left(1 - {1}/{\omega LC}\right) - \frac{R}{L}P, \quad (10)$$

$$\frac{dQ}{dt} = -\frac{3}{2L}[(V_{S,\beta}V_{R,\alpha} - V_{S,\alpha}V_{R,\beta}) + (V_{S,\beta}V_{CON,\alpha} - V_{S,\alpha}V_{CON,\beta})] + P\left(1 - {1}/{\omega LC}\right) - \frac{R}{L}Q, \quad (11)$$

in which P and Q can be subdivided into a part dependent on the line bus voltages $V_S - V_R = V_\Delta$, as well as a part dependent on the converter voltage without modifying the form of (9). The formulation of the transmission line power flow dynamics as a set of nonlinear state-space equations enables the use of DPC techniques to dampen the power flow oscillations as shown in Figs. 2(a)-(c). That is, feedback control methods such sliding mode control are then applied to (10) and (11) resulting in the damping effects observed in Figs. 2(a) – 2(c).

## V. Conclusions on Damping of Power System Oscillations

EMPT results of simulating in PSCAD [10] the implementation of the DPC design given by (10) and (11) with a static synchronous series converter (SSSC) have motivated a more detailed analysis of line compensation techniques. This paper has demonstrated the impact of line compensation through the use of an SSSC with a DPC feedback control scheme. This has been applied to a power system that is representative of a 345 kV transmission system that exhibits power system oscillations due to wind farm interactions with conventional combustion turbine power plants [9], [11].